\documentclass[twocolumn]{aastex631}
\usepackage{subfigure}

\newcommand{\cc}[1]{{\textcolor{black}{#1}}}
\begin{document}

\title{Detection of rubidium and samarium in the atmosphere of the ultra-hot Jupiter MASCARA-4b}

\correspondingauthor{Wei Wang, Meng Zhai}
\email{wangw@nao.cas.cn; mzhai@nao.cas.cn}
\author[0000-0002-0486-5007]{Zewen Jiang}
\affiliation{CAS Key Laboratory of Optical Astronomy, National Astronomical Observatories, Chinese Academy of Sciences, Datun Road A20, Beijing 100101, China}
\affiliation{School of Astronomy and Space Science, University of Chinese Academy of Sciences, Beijing 100049, China}

\author[0000-0002-9702-4441]{Wei Wang}%\thanks{E-mail: wangw@nao.cas.cn}
\affiliation{CAS Key Laboratory of Optical Astronomy, National Astronomical Observatories, Chinese Academy of Sciences, Datun Road A20, Beijing 100101, China}

\author[0000-0002-8980-945X]{Gang Zhao}
\affiliation{CAS Key Laboratory of Optical Astronomy, National Astronomical Observatories, Chinese Academy of Sciences, Datun Road A20, Beijing 100101, China}
\affiliation{School of Astronomy and Space Science, University of Chinese Academy of Sciences, Beijing 100049, China}

\author[0000-0003-1207-3787]{Meng Zhai}
\affiliation{CAS South America Center for Astronomy, National Astronomical Observatories, Chinese Academy of Sciences, Datun Road A20, Beijing 100101, China}

\author[0009-0002-4524-3076]{Yaqing Shi}
\affiliation{CAS Key Laboratory of Optical Astronomy, National Astronomical Observatories, Chinese Academy of Sciences, Datun Road A20, Beijing 100101, China}
\affiliation{School of Astronomy and Space Science, University of Chinese Academy of Sciences, Beijing 100049, China}

\author{Yujuan Liu}
\affiliation{CAS Key Laboratory of Optical Astronomy, National Astronomical Observatories, Chinese Academy of Sciences, Datun Road A20, Beijing 100101, China}

\author[0000-0003-2868-8276]{Jingkun Zhao}
\affiliation{CAS Key Laboratory of Optical Astronomy, National Astronomical Observatories, Chinese Academy of Sciences, Datun Road A20, Beijing 100101, China}

\author[0000-0002-8442-901X]{Yuqin Chen}
\affiliation{CAS Key Laboratory of Optical Astronomy, National Astronomical Observatories, Chinese Academy of Sciences, Datun Road A20, Beijing 100101, China}


\begin{abstract}

Ultra-hot Jupiters (UHJs) possess the most extreme environments among various types of exoplanets, making them ideal laboratories to study the chemical composition and kinetics properties of exoplanet atmosphere with high-resolution spectroscopy (HRS). It has the advantage of resolving the tiny Doppler shift and weak signal from exoplanet atmosphere and has helped to detect dozens of heavy elements in UHJs including KELT-9b, WASP-76b, WASP-121b. MASCARA-4b is a 2.8-day UHJ with an equilibrium temperature of $\sim2250$\,K, which is expected to contain heavy elements detectable with VLT. In this letter, we present a survey of atoms/ions in the atmosphere of the MASCARA-4b, using the two VLT/ESPRESSO transits data. Cross-correlation analyses are performed on the obtained transmission spectra at each exposure with the template spectra generated by \texttt{petitRADTRANS} for atoms/ions from element Li to U. We confirm the \cc{previous detection of Mg, Ca, Cr and Fe and report the detection of Rb, Sm, Ti+ and Ba+ with peak signal-to-noise ratios (SNRs) $>$ 5. We report a tentative detection of Sc+, with peak SNRs $\sim$6 but deviating from the estimated position}. The most interesting discovery is the first-time detection of elements Rb and Sm in an exoplanet. Rb is an alkaline element like Na and K, while Sm is the first lanthanide series element and is by far the heaviest one detected in exoplanets. Detailed modeling and acquiring more data are required to yield abundance ratios of the heavy elements and to understand better the common presence of them in UHJ's atmospheres.
\end{abstract}



\section{Introduction} \label{sec:intro}
Exoplanet science has achieved great progress in characterizing the atmospheres of exoplanets, and it becomes one of the most rapidly growing fields in astronomy. Ultra-hot Jupiters (UHJs) are a special population of Jupiter-like planets with dayside temperatures higher than 2200 K. In such conditions, most molecules are dissociated into their constituent atoms. The large atmospheric scale heights of UHJs make them ideal laboratories for the study of the physical and chemical properties of exoplanet atmospheres. Among all the powerful tools for the atmospheric characterization of exoplanets, high-resolution spectroscopy (HRS) becomes a wide-employed and efficient one, that can utilize either several strong individual lines or a few dense forests of spectral lines to search for atoms and molecules and estimate their relative abundances in the atmospheres\cc{~\citep{Snellen_2010,Brogi_2019}}. Due to much larger motion of the planets as compared to their host stars, the planetary spectra move in wavelength (or velocity) space obviously, while those from star and telluric absorption are essentially stationary, so that the atmospheric signals of an exoplanet are usually well separated from the stellar and telluric signals at high spectral resolution~\cc{\citep{Wyttenbach_2015}}. For the same reason, HRS can provide information on the relative motions of different species, and thus can be used to study the dynamics of planet atmospheres, which can not be realized with low-resolution spectroscopy in principal. 

Dozens of atoms have been detected so far using HRS in UHJs including Na, K, Ca, Mg, Ti, Fe, Co, Ca+, Fe+, Ti+, V+, Sc+, Sr+, Ni+ and Ba+. \citet{Hoeijmakers_2019} detected neutral atoms Fe, Mg, and Na, as well as ionized Sc+, Cr+, Y+, Fe+, and Ti+ in KELT-9\,b, the hottest exoplanet discovered up to today; \citet{Ehrenreich_2020} uncovered the asymmetric absorption of Fe signal in the terminators of WASP-76b; \citet{Silva_2022} reported a novel detection of Ba+ in the atmospheres of WASP-121\,b and WASP-76\,b. Ba+ is by far the heaviest element found in the exoplanet atmospheres. 

MASCARA-4\,b was identified by Multi-site All-Sky CAmeRA (MASCARA)~\citep{Talens_2017} and further confirmed by photometric observations which were taken with the Chilean–Hungarian Automated Telescope(CHAT) and radial velocity measurements which were taken with FIDEOS on the ESO 1-m telescope. In addition, follow up high-resolution spectra were taken during transit with the CTIO high-resolution spectrometer (CHIRON) instrument on the Small and Moderate Aperture Research Telescope System (SMARTS) telescope to detect the Doppler shadow of the transiting object, which further confirmed that the small object was indeed transiting the bright star. 

MASCARA-4\,b is an UHJ with a mass of $1.675\pm0.241$\,$M_{\rm Jup}$, a radius of $1.515\pm0.044$\,$R_{\rm Jup}$ and an equilibrium temperature $T_{\rm eq}$ of 2250\,$\pm\,62$\,K with an orbital period of $\sim$2.82406\,d~\citep{Dorval_2020,Zhangyapeng_2022}. Thus, it is an ideal target for atmospheric characterisation given its large atmospheric scale heights and TSM~\citep{Kempton_2018}. Recently, \citet{Zhangyapeng_2022} explored the chemical species in the atmosphere of this planet using the Echelle Spectrograph for Rocky Exoplanets and Stable Spectroscopic Observations \citep[ESPRESSO][]{Pepe_2021} mounted on the Very Large Telescope (VLT), and they resolved excess absorption about H$\alpha$, H$\beta$, Na\,{\footnotesize I} D1\&D2, Ca\,{\footnotesize II} H\&K and detected Mg, Ca, Cr, Fe, and Fe+ by the cross-correlation analysis.

Considering that quite a number of heavier elements were detected very recently in the other two UHJs WASP-76\,b and WASP-121\,b~\citep{Silva_2022} by re-analyzing the archive ESPRESSO data, it is worthy to perform a quick survey of atoms and ions in the atmosphere of MASCARA-4\,b using the HRS technique. This letter is organized as follows. We present the details of two ESPRESSO transit observations of MASCARA-4\,b and data reduction in Section 2. The data analysis including transmission spectrum construction and atmospheric cross-correlation analysis are given in section 3. The final results and some discussions are presented in section 4.

\section{OBSERVATIONS AND DATA REDUCTION}\label{sec:telluric correction}

Two transits of MASCARA-4\,b were observed in 2020 February 13 and 2020 March 1 with ESPRESSO under the ESO programs 0104.C-0605 (PI:WYTTENBACH). ESPRESSO is a fibre-fed ultra-stable échelle high-resolution spectrograph, mounted at the 8.2m Very Large Telescope at European Southern Observatory in Cerro Paranal, Chile \citep{Pepe_2021}. This observation was carried out using the single-UT HR21 mode with a spectral resolving power $R\sim140000$ and wavelength coverage of $380-788$\,nm. For the transit in 2020 Feb 13 (hereafter T1), a total of 96 spectra were obtained with 39 taken in transit and 57 out of transit with orbital phase $\Phi$ from -0.053 to 0.087. For the transit in 2020 Mar 1 (hereafter T2), a total of 85 spectra were achieved with 25 taken in transit and 60 out of transit, covering $\Phi$ from -0.07 to 0.066. Details of the two transits are summarized in Table~\ref{obs}. We used the 1D spectra from the ESO data archive, which were processed by the ESPRESSO reduction pipeline, version 2.2.1. Bias, dark, flat and bad pixels were taken into account when processing the 2D images, from which 1D spectra can be sky-subtracted and extracted. Then all the échelle orders were wavelength calibrated and stitched into a combined 1D spectra.

%%%%%%%%%%%%%%%%%%%%%%%%%%%%%%%%%%%%%%%%%%%%%%%%%%%%%
   \begin{table*}
      \centering
      \caption{Summary of the MASCARA-4b transit observations.}
         \label{obs}
         \begin{tabular}{lcccccccc}
            \hline
            \hline
            \noalign{\smallskip}
            & Date &\multicolumn{3}{c}{Number of spectra} & Exp. time & Airmass range & Mean S/N & Program ID\\
            \cline{3-5}
            & (UT) &  Total & In-transit & Out-of-transit & (s) &(start$-$T$_{\rm c}-$end)  & (@580\,nm) &   \\
            \hline
            T1 & 2020-02-13 & 96 & 39 & 57 &360 & $1.99-1.33-1.91$ & $\sim$211 & 0104.C-0605  \\
            T2 & 2020-03-01 & 85 & 25 & 60 &300 & $1.72-1.34-2.46$ & $\sim$207 & 0104.D-0605 \\
            \hline
         \end{tabular}
         \newline
     \end{table*}
%%%%%%%%%%%%%%%%%%%%%%%%%%%%%%%%%%%%%%%%%%%%%%%%%%%%%
We corrected the telluric absorption hidden in the spectra using the ESO software \texttt{Molecfit} version 1.5.7~\citep{Smette_2015,Kausch_2015}. \texttt{Molecfit} is a tool for modelling telluric lines using the state-of-the-art radiative transfer modelling of the Earth’s atmosphere, together with a comprehensive database of molecular parameters~\citep{Smette_2015}. It has been widely used recently in the HRS studies of exoplanet atmosphere\citep[e.g.,][]{Allart_2017, Seidel_2020, Kirk_2020} to best remove telluric contamination. 

%\citet{Allart_2017} removed telluric features in the spectrum using \texttt{Molecfit} in order to search for water vapor of HD 189733b; ~\citet{Seidel_2020} corrected the telluric lines with \texttt{Molecfit}  and detected \ion{Na}{I} using VLT/ESPRESSO at the edge of the Neptune desert for WASP-166b, an extreme-puffy planet. ~\citet{Kirk_2020} found peak excess absorption of 7.26\% $\pm$ 0.24\% centered on the \ion{He}{I} triplet at 10833\AA\, using Keck II/NIRSPEC after eliminating the influence of telluric contamination with Molecfit.

\begin{figure}
	\includegraphics[width=\columnwidth]{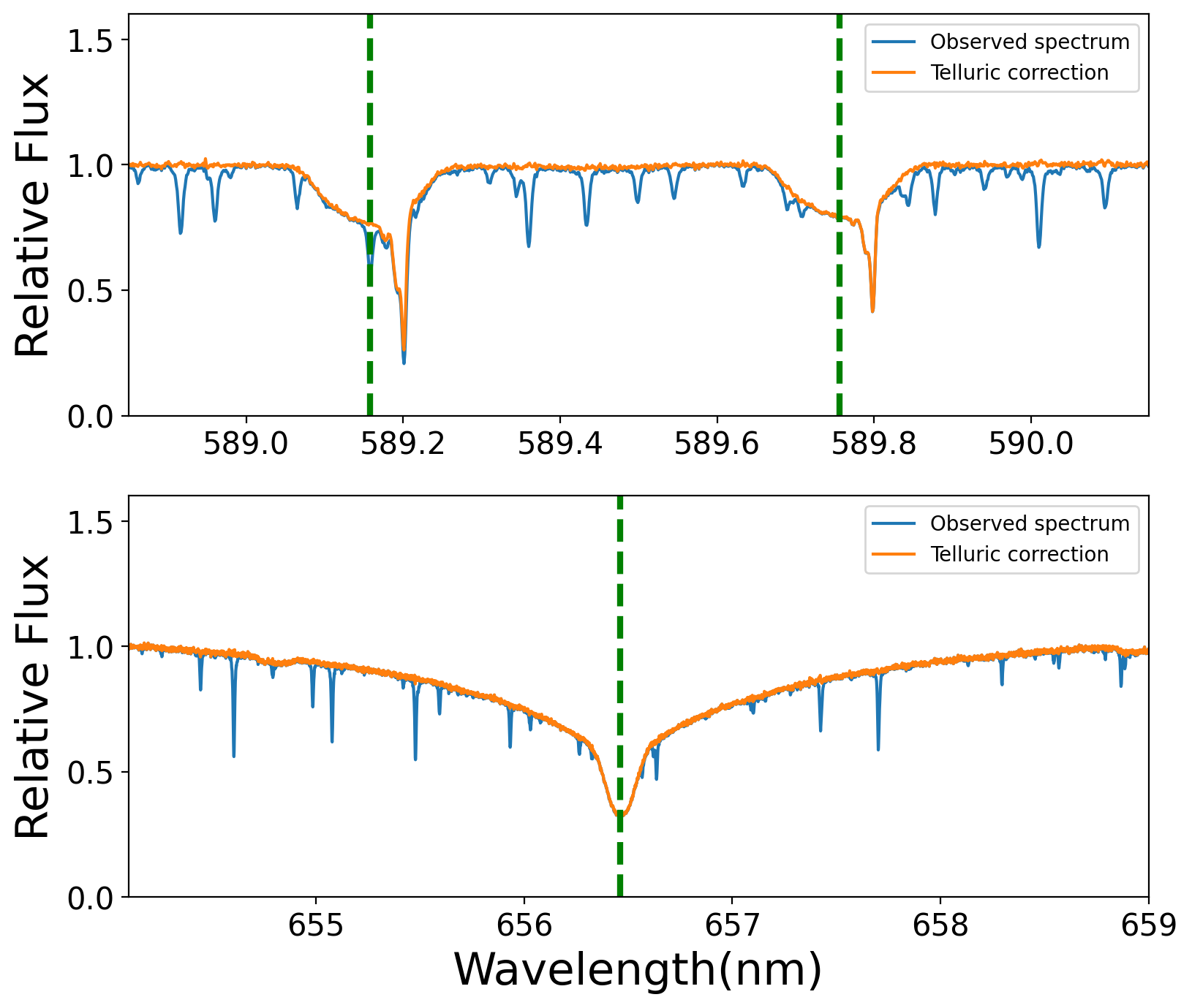}
    \caption{An example of telluric correction using \texttt{Molecfit} in the observed spectrum of MASCARA-4\,b taken in T1 for Na D1 $\&$ D2 (\textit{Top}) and H$\alpha$ (\textit{Bottom}). The observed spectra are shown in blue, while the spectra after telluric correction are shown in orange. The dashed green vertical lines represent the static positions of the atomic lines in vacuum.}
    \label{fig:telluric_correction_figure}
\end{figure}

We adopted the input parameters for \texttt{Molecfit} as described in ~\citet{Allart_2017}. To achieve a good estimate and removal of telluric contamination, the fitting regions were manually chosen by visual inspection and comparisons of the observed spectra and the model telluric atmospheric transmission spectra. Note that the obtained ESPRESSO spectra are given in the solar system barycentric rest frame, which were shifted to terrestrial rest frame to fulfill the \texttt{Molecfit} requirement. As an example, we show in Fig.~\ref{fig:telluric_correction_figure} comparisons of the observed spectrum taken on T1 and its corresponding contamination-corrected spectrum around the Na\,{\footnotesize I} D1$\&$D2 doublet region (Top panel) and H$\alpha$ spectral line region (Bottom panel), illustrating the goodness of our telluric correction. 
%%%%%%%%%%%%%%%%%%%%%%%%%%%%%%%%%%%%%%%%%%%%%
% First figure
   \begin{table}
   \centering
   
      \caption{Parameters of the MASCARA-4 system. }
         \label{system} 
           \begin{tabular}{lll}
            \hline
            \hline
            Description &Symbol & Value \\
            \hline
            Stellar Parameters & & \\
            \hline
            $V$ magnitude& $m_{\rm v}$    & $8.19\pm0.01$\,mag    \\
            Effective temp.& $T_{\rm eff}$& $7800\pm200$\,K   \\          
            Surface gravity&  log\,$g_\star$  & $4.10\pm0.05$ cgs\\    
            Metallicity& [Fe/H] &  $\sim0.00$\,dex   \\
            Stellar mass&  $M_\star$  & $1.75\pm0.05\,M_\odot$   \\
            Stellar radius &$R_\star$ & $1.92\pm0.11\,R_\odot$   \\
            Projected spin&V\rm sin$i_{\star}$ & $45.66_{-0.9}^{+1.13}\,\rm {km\,s^{-1}}$\\
            \noalign{\smallskip}
            \hline
		    Planet Parameter & & \\
            \hline
            \noalign{\smallskip}
            Planet mass$^{1}$   & $M_{\rm p}$ & $1.675\pm0.241\,M_{\rm Jup}$   \\  
            Planet radius$^{1}$ & $R_{\rm p}$ & $1.515\pm0.044\,R_{\rm Jup}$ \\           
            Planet density$^{1}$ &$\rho$  & $0.481_{-0.079}^{+0.085}$\,g\,cm$^{-3}$ \\          
            Eequilibrium temp$^{1}$& $T_{\rm eq}$ & $2250\pm62$\,K  \\  
            Radius ratio$^{1}$& $R_{\rm p}/R_\star$ & $0.08_{-0.002}^{+0.004}$\\
            \noalign{\smallskip}
            \hline
            Orbit Parameters& &\\
            \hline
            Epoch-BJD& $T_{\rm c}$ & $2458505.817\pm0.003$ \\ 
            Period & $P$  & $2.82406\pm0.00003$\,d    \\ 
            Transit duration& $T_{14}$ & $0.165\pm0.006$\,d \\ 
            Semi-major axis & $a$ &  $0.047\pm0.004$\,AU   \\ 
            Inclination &$i$ & $88.5\pm0.01$\,deg \\   
            Semi-amplitude velocity$^{1}$&$K_{\rm p}$&$182\pm5$\,\rm km$\,\rm s^{-1}$\\
            \hline
            \noalign{\smallskip}
         \end{tabular}
        \textbf{Notes}:$(1)$~\citet{Zhangyapeng_2022}. All other parameters refer to~\citet{Dorval_2020}.
   \end{table}
%%%%%%%%%%%%%%%%%%%%%%%%%%%%%%%%%%%%%%%%%%%%%%%%%%

\section{DATA ANALYSIS}
\subsection{Transmission spectrum construction}
\label{sec:Transmission_spectra}
The signal of an exoplanet atmosphere is encoded in transmission spectra of the planet. Following the previous successes in HRS studies, we adopted the similar method outlined in \citet{Wyttenbach_2015} and \citet{Casasayas-Barris_2019} to extract transmission spectra for each exposure. After correcting telluric contamination as described in Section~\ref{sec:telluric correction}, all the observed spectra were normalized to their continuum level with the \texttt{iSpec} package ~\citep{Blanco-Cuaresma_2014,Blanco-Cuaresma_2019}. We also applied a sigma-clipping rejection algorithm on the normalized spectra and replaced the cosmic ray hits with the mean value of all the other spectra at each wavelength~\citep{Allart_2017}. Then we shifted the spectra to the stellar rest frame, and constructed a master-out spectrum by calculating the average of the aligned out-of-transit spectra with their mean SNR as weights. The master-out spectrum is should be purely stellar origin, and is thus used to divide each individual spectrum in order to remove stellar light contribution so that the absorption signal from the planetary atmosphere may stand out~\citep{Stangret_2021}. The resulted ``divide-out'' residual spectra constitute the 2D transmission spectrum map.

\subsection{\cc{Survey of atoms and  molecules using the cross-correlation technique}}
\label{sec:survey atom and molecule} 

\subsubsection{\cc{Template spectra}}
\label{sec:atmosphere template}
\citet{Zhangyapeng_2022} reported excess absorption by H$\alpha$, H$\beta$, Na\,{\footnotesize I} D1$\&$D2, Ca\,{\footnotesize II} H\&K, and strong lines of Mg, Fe and Fe+, and detected the species Mg, Ca, Cr, Fe and Fe+ using the cross-correlation technique. Given its high equilibrium temperature, we suspected that this planet might also bear heavier species, as found in the atmospheres KELT-9\,b, WASP-121\,b and WASP-76\,b~\citep{Hoeijmakers_2019, Ehrenreich_2020, Silva_2022}. Therefore, we performed a thorough survey of the atoms/ions from Li to U in the atmosphere of MASCARA-4\,b in the obtained high-R transmission spectra using the cross-correlation function (CCF) technique. That is to calculate the CCFs of the obtained transmission spectra and the template spectra, following the pioneer works~\citep{Snellen_2010,Hoeijmakers_2019}. This technique can fully utilize all the spectral signals from multiple lines of a certain species to enhance the detection capability. 

The template spectra for a given species or a combination of multiple species in planet atmosphere can be computed by the python package \texttt{petitRADTRANS}~\citep[][pRT in short]{P_Molliere_2019}. This package is built for the spectral characterization of exoplanet atmosphere and has been used successfully in many atmospheric studies~\citep[e.g. ][]{PMolliere_2020,Landman_2021,Casasayas_Barris_2022}, which requires inputs of parameters of the planet system, a pressure-temperature (TP), and opacities for each species of interests. The planetary and stellar parameters that are required are all included in Table~\ref{system}. An isothermal pressure-temperature (TP) profile with a temperature of 2500\,K is adopted and a solar abundance is assumed for the determination of the volume mixing ratio (VMR) of various species. However, for some species it is not that straightforward to generate their model spectra, because of the lack of corresponding opacities in the pRT database. This prevent a free exploration of any species in the planet atmospheres. Fortunately, the DACE database provides pre-calculated opacities for many neutral and ionized atoms from the element Li to element U. There opacities are generated using the method presented in ~\citet{Grimm_2015}, at a wide temperature range of 2000\,K to 6000\,K but at a constant pressure of $10^{-8}$\,bar. \citet{Kesseli_2022} carried out a comparison study between the pRT pre-computed opacities with those from the DACE opacities, and they concluded that no significant difference in signals should be expected. \cc{For comparisons, we perform additional CCF analysis using the DACE opacities for the species with pRT pre-computed opacities available, i.e., Mg, Ca, Cr and Fe. We confirm the conclusion drawn by \citet{Kesseli_2022} that both templates yield consistent results, particularly the detection S/Ns and signal's velocity, as shown in Table~\ref{Gaussian fit}}

\subsubsection{\cc{Cross-correlation analysis}}
\label{sec:ccf}
\begin{figure}
	\includegraphics[width=\columnwidth]{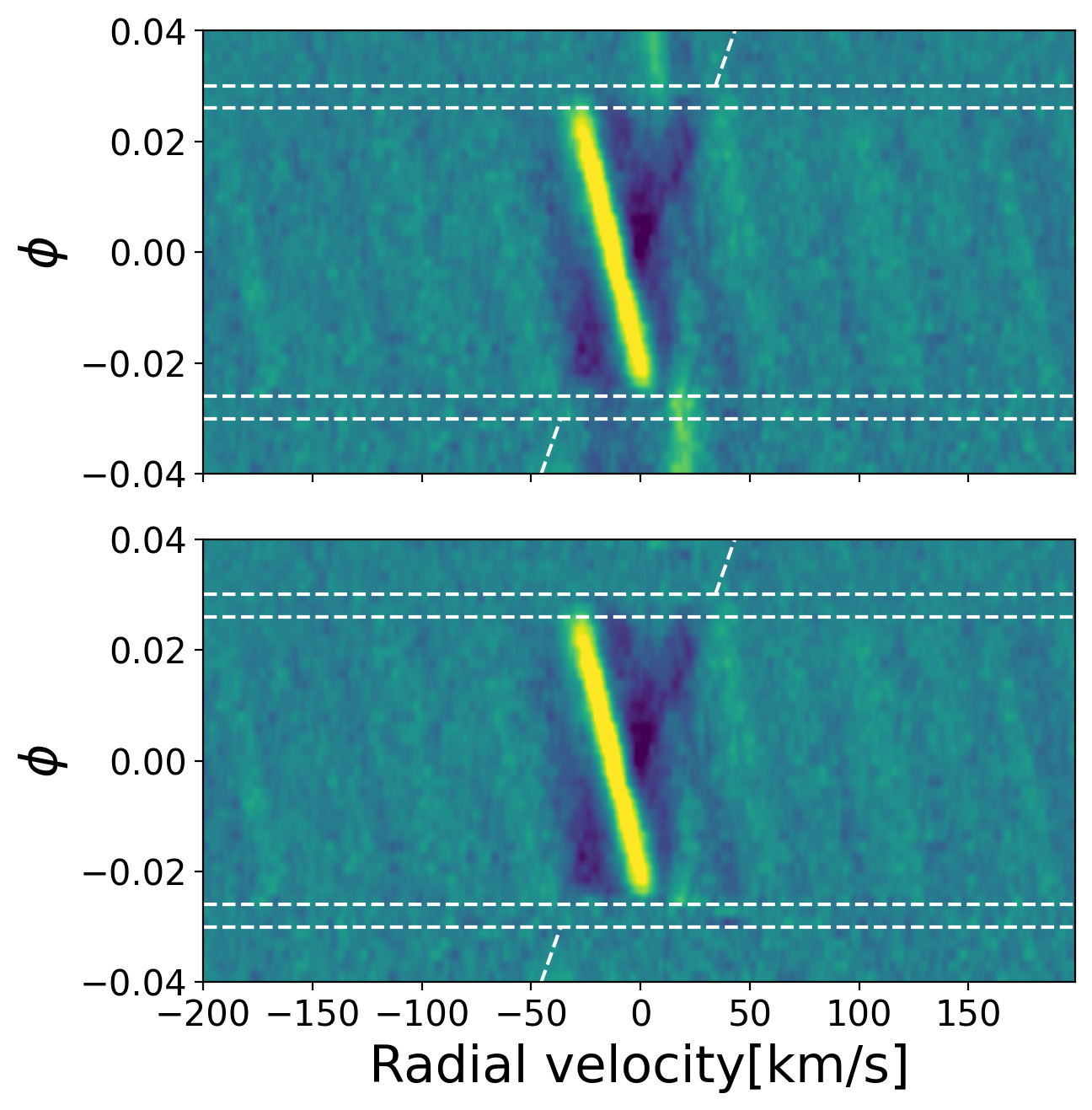}
    \caption{\cc{The combined CCF Map at different orbital phase before (\textit{top panel}) and after (\textit{bottom panel}) the correction of stellar pulsation, respectively. The Map is combined with a step size of 0.002 in orbital phase. In each panel,} the white dotted lines mark the four contacts during the transit, the inclined white lines indicate the expected trace of signal from the planet, \cc{and the bright yellow stripe shows the RM +CLV signal. The out-of-transit steaks in the top panel arise from stellar pulsation.}}
    \label{fig:stellar pulsation}
\end{figure}

The generated template spectra are then convolved with a Gaussian function to match the resolution of spectra obtained with ESPRESSO. \cc{Then, the CCF coefficients between the template spectra and the stellar rest frame residual spectra are calculated using} the following formula:
\begin{equation} \label{ccf}
\centering
	c(v,t)= \frac{\sum_i^N x_i(t) T_i(v)}{\sum_i^N T_i(v)},
\end{equation}

where $T_i(v)$ is the template shifted to a RV of $v$, $x_i(t)$ is the transmission spectrum at time $t$, $c(v,t)$ is a 2-dimensional matrix dependent on $t$ and $v$. The CCFs are calculated for each residual transmission spectrum and template spectrum, with the latter moving in the RV range of [$-100,100$]\,km\,s$^{-1}$ with step of 1\,km\,s$^{-1}$, resulting a 2D residual CCF map as shown in Fig.~\ref{fig:stellar pulsation}. If the targeted species exists, the signal shall appear as a dark trace along the expected moving direction of the planet in the CCF map, and show at the position of the estimated orbital velocity $K_{\rm p}$ and system velocity $V_{\rm sys}$ in the $K_{\rm p}-\Delta V_{\rm sys}$ map~\citep{Hoeijmakers_2019}. 
%%%%%%%%%%%%%%%%%%%%%%%%%%%%%%%%%%%%%%%%%%%%%%%%%%%%%
%%%%%%%%%%%%%%%%%%%%%%%%%%%%%%%%%%%%%%%%%%%%%%%%%%%%%

To achieve higher signal-to-noise ratios (SNRs) for the purpose of gaining higher detection significance, the CCF maps obtained for T1 \& T2 were combined into one CCF map. The CCF spectra in one night are firstly interpolated at a uniform phase grid with a constant step of 0.002, such that the two nights CCF maps are aligned in phase space. Then for each phase, the two interpolated CCF spectra from T1 \& T2 are averaged to be the new combined CCF spectrum. Repeating this procedure for each phase, one can obtain the combined CCF map. An example is shown in Figure~\ref{fig:stellar pulsation} for the atom Fe. The dark shadow in between the inclined dashed lines represents the absorption of the planet's atmospheric signal, while the bright stripe is resulted from the center-to-limb (CLV) variation~\citep{Yanfei_2017}) and the Rossiter-McLaughlin (RM) effect~\citep{Rossiter_1924} and \citep{McLaughlin_1924}, due to the non-uniform of the brightness in the stellar discs and the differences in the projected RVs at the different distance to the stellar centers. 

\subsubsection{\cc{Stellar pulsations}}
\label{sec:stellar-pulsations}
\cc{As shown in the top panel of Figure~\ref{fig:stellar pulsation}, two bright streaks are visible in the out-of-transit region, which should be caused by the stellar pulsations. Such patterns are also noted in the residual CCF Map of \cite{Zhangyapeng_2022} and need to be carefully removed. We follow the method presented in \citet{Wyttenbach_2020} and \citet{Zhangyapeng_2022} to empirically mitigated the ripples, assuming that the stellar pulsation feature in each row of the CCF map can be approximately represented by a 1D Gaussian function with constant radial velocity and can be reasonably modelled. The out-of-transit residual CCFs before ingress and after egress are co-added respectively, with the strongest peak fitted by a Gaussian profile. All the individual out-of-transit CCFs are subtracted with the best-fit Gaussian component. We iterate this fitting process seven times until the two streaks are barely visible in the CCF map. As shown in the bottom panel of Figure~\ref{fig:stellar pulsation}, the out-of-transit pulsation signal is well mitigated, although there still some weak structure remaining in the in-transit region.
}
%%%%%%%%%%%%%%%%%%%%%%%%%%%%%%%%%%%%%%%%%%%%%%%%%%%%%
\begin{figure}
	\includegraphics[width=9cm, height=4.8cm]{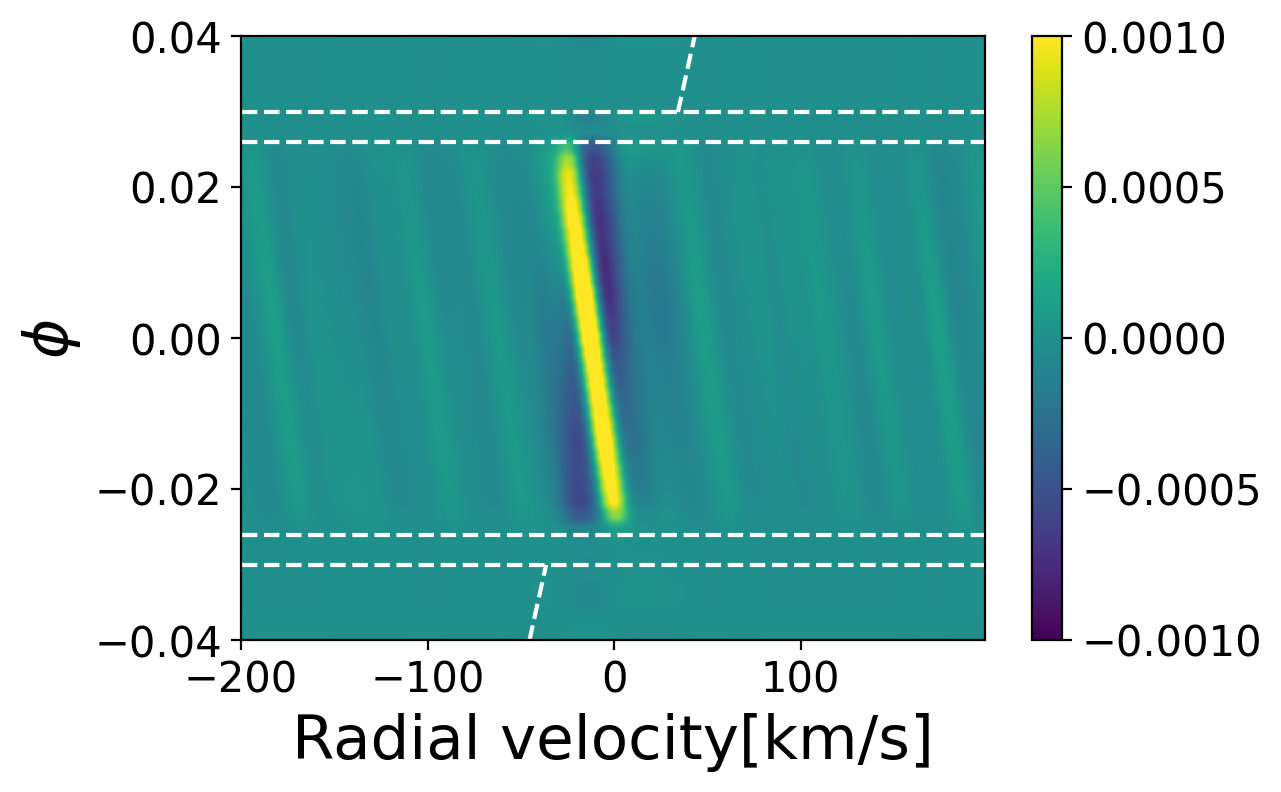}
    \caption{\cc{The RM+CLV model for Fe generated by cross-correlation between residual spectra created by synthetic spectrum using \texttt{SME} and the template spectra calculated by DACE, which is used to eliminate the influence of the variation of stellar line profile during transit.}}
    \label{fig:RM+CLV model}
\end{figure}
%%%%%%%%%%%%%%%%%%%%%%%%%%%%%%%%%%%%%%%%%%%%%%%%%%%%%
\subsubsection{\cc{Removal of the RM and CLV Effect}}
\label{sec:remove rm and clv effect}

\cc{The deformation of the stellar line profile caused by CLV and RM effects produce strong signals in the CCF maps (the yellowish bright strip), which mixes with the planet signal and prevents a reliable study on the latter signal. To disentangle the stellar signal from the planetary signal, we follow the method used in \citet{Yan_2018}, \citet{Chen_2020} and \citet{Casasayas_Barris_2019} to model the stellar spectra at different transit positions. We use the \texttt{Spectroscopy Made Easy} tool (SME,\citealt{Valenti_1996}) to compute the theoretical stellar spectra at 21 different limb-darkening angles ($\mu$) using the MARCS and VALD3 line lists~\citep{Ryabchikova_2015} and employ solar abundance and local thermodynamical equilibrium (LTE) for the calculation of stellar spectra. The stellar disk is divided into small elements with size of $0.01\,R_\star\times0.01\,R_\star$, and each of them owns specific parameters including $v\,{\rm sin}\, i_{\star}$, $\mu$, and $\theta$~( the angle between the normal to each element and the line of sight). The relative position of the planet to the stellar disk is calculated assuming a uniform velocity during the transit. Then, the synthetic spectrum during transit is calculated by summing up the spectra of all the surface elements excluding those obscured by the planet.}

\cc{The next step is to simulate the RM+CLV effects and correct them. We construct the master out-of-transit spectrum of the synthetic spectrum and divided each synthetic spectrum by it and then cross-correlated with the generated template spectra for each species. The obtained RM+CLV 2D CCF map is to be used as a proxy as shown in the Figure~\ref{fig:RM+CLV model}, which will be re-scaled to match the observation and subtracted. After the correction, the influence of RM+CLV effect are largely reduced.}

%ions of Na, Mg, Ca, Cr, Fe, Rb, Ti+, Ba+, Mn, Sc+,Sm, Cr+,Y+,Zr+
%%%%%%%%%%%%%%%%%%%%%%%%%%%%%%%%%%%%%%%%%%%%%%%%%%%%%
   \begin{table*}
      \caption{Summary of the detection \cc{derived by DACE and \texttt{pRT}}. In the last column, ``Confirmative detection" means the species were detected by previous work, and is confirmed in this work; ``First detection" means this species is detected for the first time in exoplanets; ``New discovery" means this species have been detected in other exoplanets but new in MASCARA-4b, and this work present the first detection of this species in MASCARA-4b.}
         \label{Gaussian fit}
         \centering
         \begin{tabular}{lccccccc}
            \hline
            \hline
            \noalign{\smallskip}
            Species & Template &Peak & Amplitude & $\rm K_{\rm p}$ & $V_{\rm shifted}$ & FWHM & Remarks\\
            &Type&S/N &ppm&  \rm km\,s$^{-1}$ & \rm km\,s$^{-1}$&\rm km\,s$^{-1}$ \\
            Mg &pRT&6.1 &1387.82\,$\pm\,108.0$  & 147.59\,$\pm\,28.06$&-0.8\,$\pm\,1.91$& 11.63\,$\pm\,0.34$ & Confirmative detection\\
               &DACE&7.3 &2632.80\,$\pm\,176.0$  & 163.59\,$\pm\,24.94$&0.6\,$\pm\,1.88$& 13.68\,$\pm\,0.33$ &\\
            Ca &pRT&5.2 &162.86\,$\pm\,18.4$  & 198.59\,$\pm\,22.20$&-11.8\,$\pm\,2.88$&14.96\,$\pm\,0.47$& Confirmative detection\\
               &DACE&5.2 &376.05\,$\pm\,36.7$  & 186.59\,$\pm\,27.48$&-10.3\,$\pm\,3.48$&18.17\,$\pm\,0.48$&\\
            Cr &pRT&7.4 &2084.84\,$\pm\,183.0$  & 201.59\,$\pm\,18.18$&-2.2\,$\pm\, 2.34$& 17.24\,$\pm\,0.52$& Confirmative detection\\
               &DACE&6.5 &1108.44\,$\pm\,80.50$  & 223.59\,$\pm\,30.00$&-11.6\,$\pm\, 2.95$& 19.00\,$\pm\,0.46$&\\
            Fe &pRT&14.0 &6096.99\,$\pm\,296.0$  & 181.59\,$\pm\,14.02$&-1.1\,$\pm\, 0.92$& 12.87\,$\pm\,0.24$& Confirmative detection\\
               &DACE&14.9 &5827.07\,$\pm\,265.0$  & 195.59\,$\pm\,12.72$&-2.2\,$\pm\, 0.92$& 13.66\,$\pm\,0.23$&\\

            \noalign{\smallskip}
            Rb &DACE&5.9 &145.03\,$\pm\,12.80$  & 183.59\,$\pm\,24.14$&-3.8\,$\pm\, 3.28$& 19.32\,$\pm\,0.62$ & First detection\\
            Sm &DACE&5.1 &170.66\,$\pm\,17.2$  & 205.59\,$\pm\,23.75$&-3.4\,$\pm\,3.53$& 17.84\,$\pm\,0.52$& First detection\\
                        \noalign{\smallskip}
            Ti+ &DACE&7.9 &3059.07\,$\pm\,233.0$  & 175.59\,$\pm\,17.98$&-1.1\,$\pm\,1.53$& 11.67\,$\pm\,0.35$& New discovery\\
            Ba+ &DACE&5.5 &61.52\,$\pm\,4.64$  & 169.59\,$\pm\,55.40$&-1.4\,$\pm\, 7.49$& 41.35\,$\pm\,0.83$& New discovery\\
                        \noalign{\smallskip}
            Sc+ &DACE&6.4 &622.73\,$\pm\,35.9$  & 219.59\,$\pm\,32.75$&-22.3\,$\pm\, 3.89$& 25.01\,$\pm\,0.42$&Tentative detection\\
            \hline
         \end{tabular}
        %\tablenotes{These parameters are derived by Gaussian fit using the python module lmfit. The uncertainties of parameters are calculated adopting the method as described in \citet{Kesseli_2022}.}
     \end{table*}
%%%%%%%%%%%%%%%%%%%%%%%%%%%%%%%%%%%%%%%%%%%%%%%%%%%%%

\section{RESULTS AND DISCUSSION}
\begin{figure*}
    \centering
    \subfigure{
    \begin{minipage}[t]{1\linewidth}
    \centering
    \includegraphics[width=17cm]{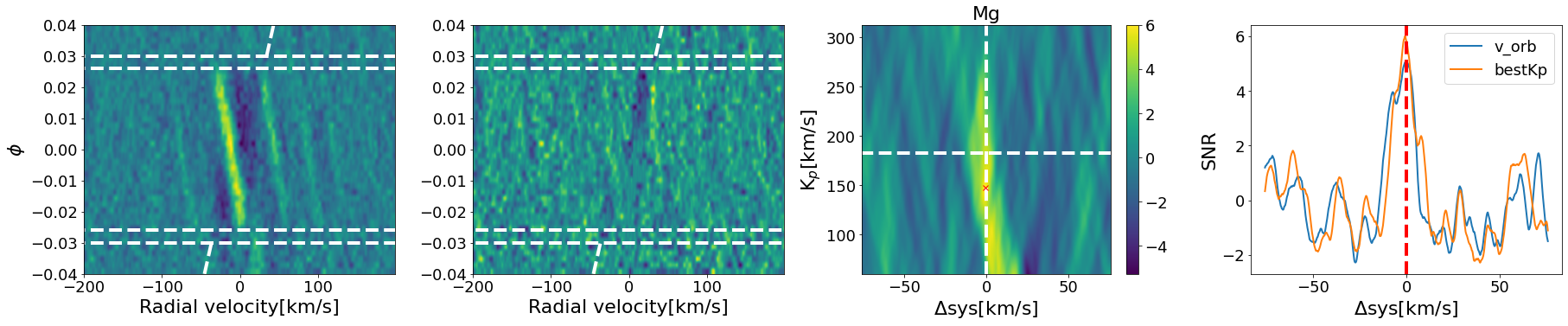}  
    \end{minipage}
    }
    
    \subfigure{
    \begin{minipage}[t]{1\linewidth}
    \centering
    \includegraphics[width=17cm]{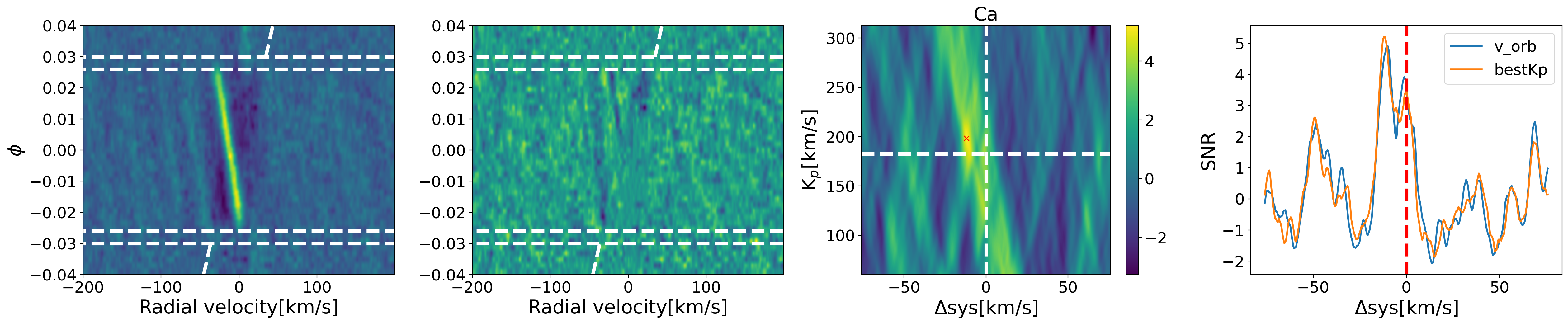}  
    \end{minipage}
    }
    
    \subfigure{
    \begin{minipage}[t]{1\linewidth}
    \centering
    \includegraphics[width=17cm]{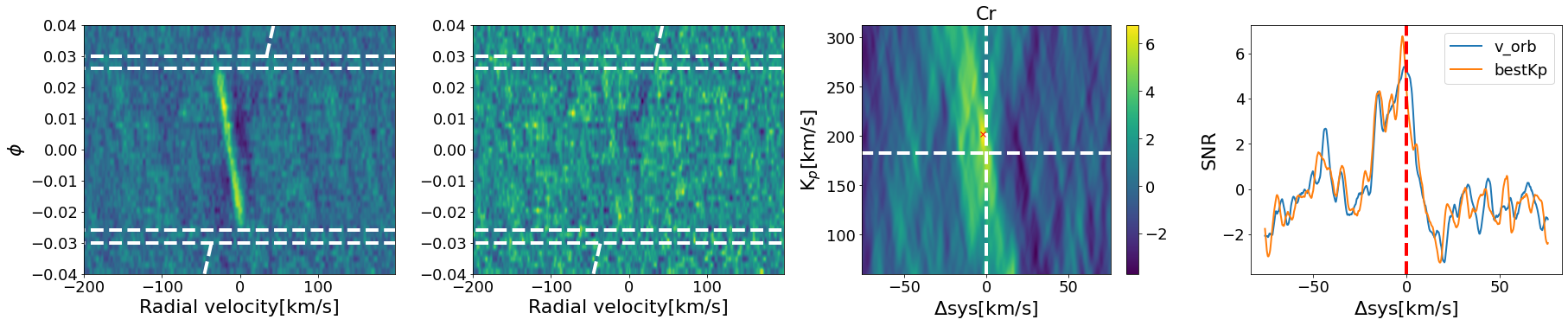}  
    \end{minipage}
    }

    \subfigure{
    \begin{minipage}[t]{1\linewidth}
    \centering
    \includegraphics[width=17cm]{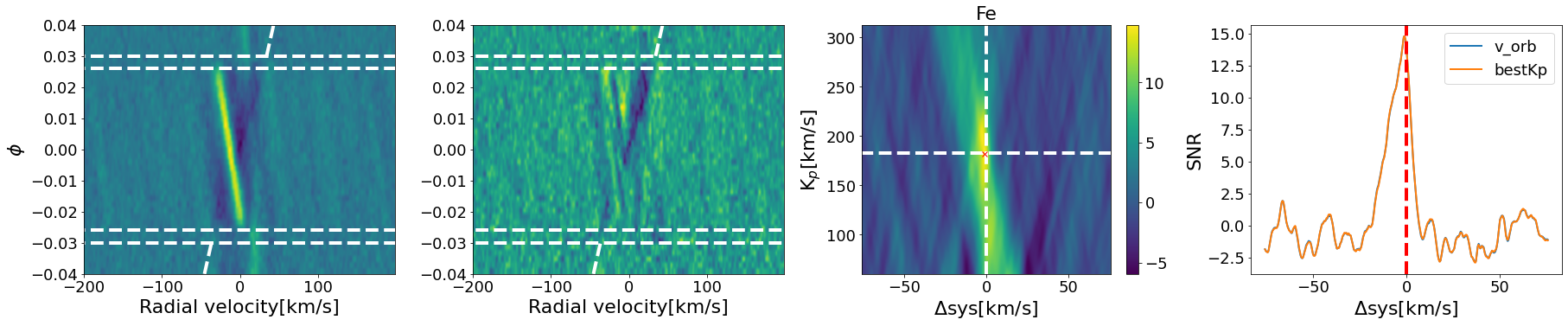}  
    \end{minipage}
    }
    
\caption{\cc{\emph{First panels}: The 2D CCF maps of Mg, Ca, Cr and Fe with CLV+RM effects uncorrected using \texttt{pRT} opacity. The white dotted lines mark the beginning and ending positions of the transit and the inclined white lines indicate the expected trace of signal from the planet. \emph{Second panels}: Same as \emph{the first panels} but with CLV+RM effects corrected. \emph{Third panels}: the $K_{\rm p}$-$\Delta V_{\rm sys}$ maps in the range of $50\sim300$\, km\,s$^{-1}$. In each panel, the planet signal is expected to appear around the intersection of the two white dotted lines, while the red crosses marks the position with maximum SNR. \emph{Fourth panels}: the SNR plots at the expected $K_{\rm p}$ position in blue and at the Max-SNR corresponding $K_{\rm p}$.}}

    \label{fig:ccf_result_petit_1}
\end{figure*}

\begin{figure*}
    \centering
    \subfigure{
    \begin{minipage}[t]{1\linewidth}
    \centering
    \includegraphics[width=17cm]{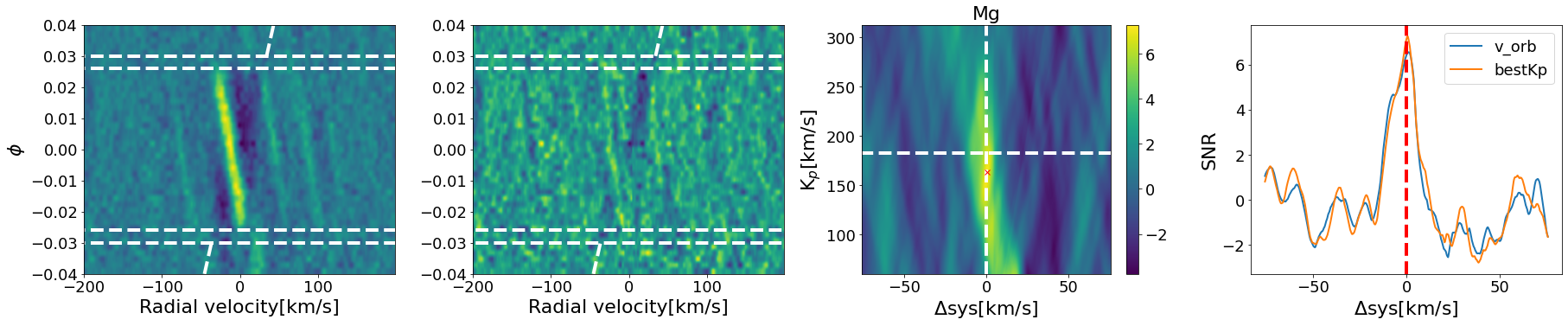}  
    \end{minipage}
    }
    
    \subfigure{
    \begin{minipage}[t]{1\linewidth}
    \centering
    \includegraphics[width=17cm]{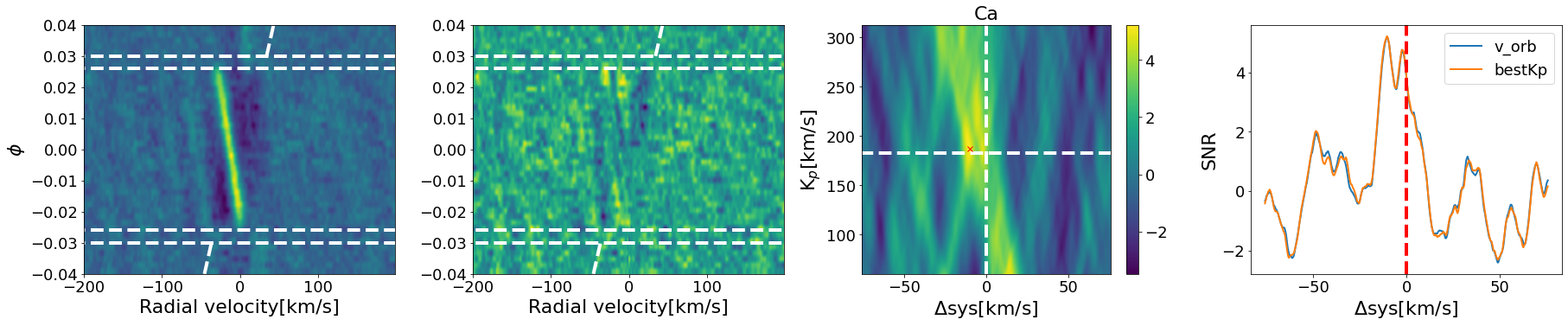} 
    \end{minipage}
    }
    
    \subfigure{
    \begin{minipage}[t]{1\linewidth}
    \centering
    \includegraphics[width=17cm]{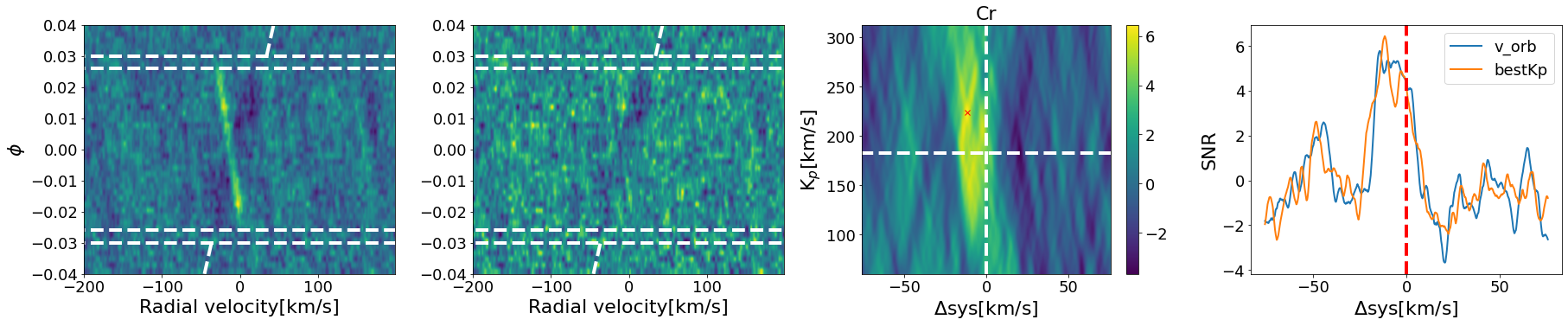}  
    \end{minipage}
    }

    \subfigure{
    \begin{minipage}[t]{1\linewidth}
    \centering
    \includegraphics[width=17cm]{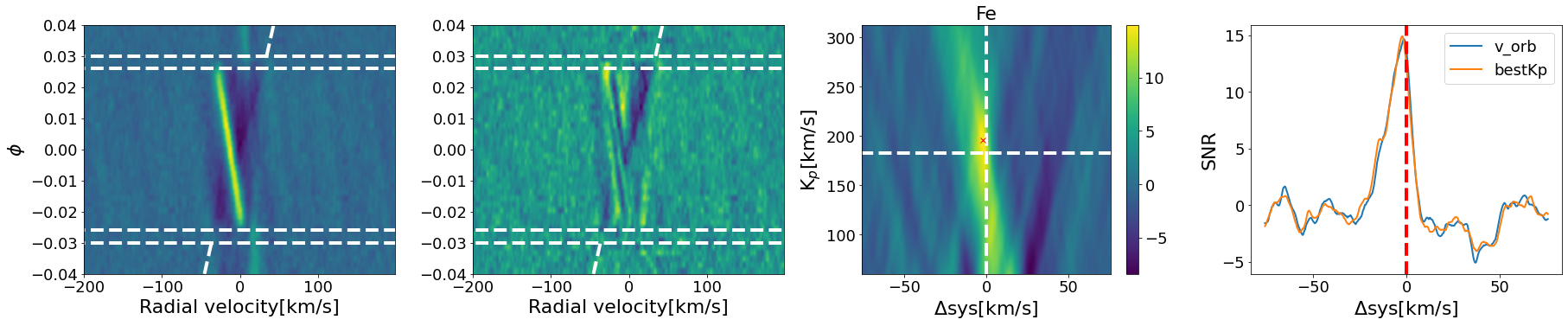}  
    \end{minipage}
    }

\caption{Same as Fig~\ref{fig:ccf_result_petit_1}: but using DACE opacity to obtain CCFs map for Mg, Ca, Cr and Fe.}
    \label{fig:ccf_result_dace_1}
\end{figure*}

The obtained transmission spectra are checked visually around the strong individual absorption lines. We confirm that H$\alpha$, H$\beta$, Na\,{\footnotesize I} doublet, Ca\,{\footnotesize II} doublet, Mg\,{\footnotesize I}, Fe\,{\footnotesize I} and Fe\,{\footnotesize II} do show additional absorption, as claimed by \citet{Zhangyapeng_2022}. The cross-correlation analysis are performed \cc{respectively} for all the species with available opacities from Li to U \cc{generated by DACE, and those covered by \texttt{pRT} using its own database}. In Table~\ref{Gaussian fit}, we list all the species with absorption features detected with peak S/Ns $>5$. \cc{As discussed in Section~\ref{sec:atmosphere template}, the results using different soures of opacities are consistent with each other}. We confirm from Fig.~\ref{fig:ccf_result_petit_1} \cc{and Fig.~\ref{fig:ccf_result_dace_1} } clear detection of Mg, Ca, Cr and Fe firstly reported by \citet{Zhangyapeng_2022}. \cc{The detection significance, i.e., the S/N values, of Mg, Ca and Cr in this work are well consistent with those in \citet{Zhangyapeng_2022}. However our detection S/N of Fe in the CCF map is obviously smaller than that in \citet{Zhangyapeng_2022}, but is consistent with their obtained values using individual strong lines (cf.~their Table 3.)}

In Figures~\ref{fig:ccf_result_petit_1} \& \ref{fig:ccf_result_dace_1}, each row contains \cc{4 panels} figures for one species. The first \cc{and second panels show the 2D CCF maps for each species before and after the CLV+RM correction, respectively}, where the inclined white dashed lines indicate the expected track of planet signal. Note that the left-inclined \cc{bright yellow stripe} is \cc{the CLV+RM signals}; the third panel is the $K_{\rm p}-\Delta V_{\rm sys}$ map, created by co-adding CCFs assuming a range of $K_{\rm p}$ and $\Delta V_{\rm sys}$, where $K_{\rm p}$ is the Keplerian speed of the planet and is estimated to be $\sim182$\,km\,s$^{-1}$ by \citet{Zhangyapeng_2022} while $\Delta V_{\rm sys}$ is the velocity relative to the planet rest frame. In such maps, signal from the planet should be the strongest around the expected $K_{\rm p}$ and $\Delta V_{\rm sys}$, i.e., at the intersection of white dashed lines. The red cross marks the position with maximum SNR, which do not always coincide well with the intersection point. Gaussian fits using the python module \texttt{lmfit} are applied on the horizontal and vertical slices passing the red cross to the determine the peak positions in $\Delta V_{\rm sys}$ and $K_{\rm p}$, respectively, as well as the FWHM of the detected signals. The best-fit parameters are listed in Table~\ref{Gaussian fit}, together with their uncertainties calculated adopting the method described in \citet{Kesseli_2022}. The \cc{fourth} panel shows the SNR profile of the corresponding species against different $\Delta V_{\rm sys}$, either at the expected $K_{\rm p}$ (the blue line), or at the $K_{\rm p}$ with maximum SNR (the orange line), respectively.

\begin{figure*}
    \centering
    
    \subfigure{
    \begin{minipage}[t]{1\linewidth}
    \centering
    \includegraphics[width=17cm]{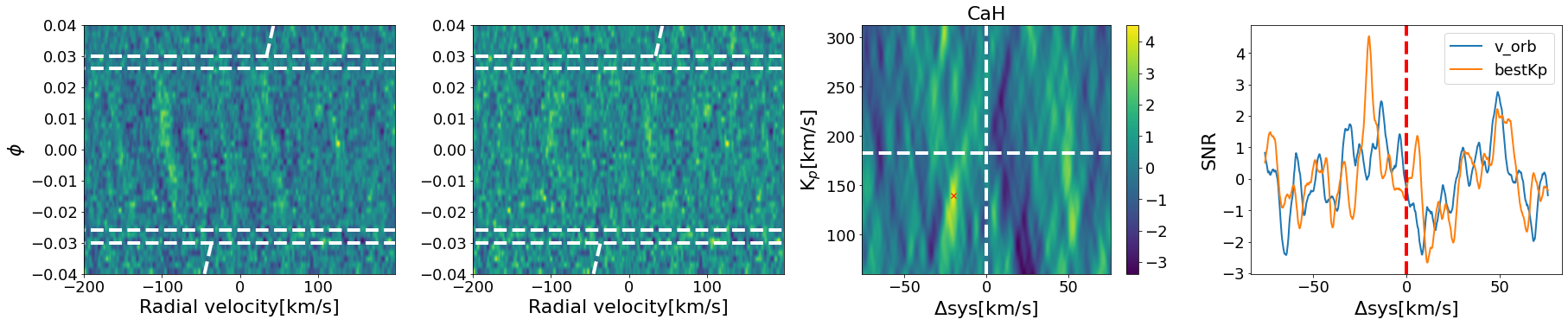}  
    \end{minipage}
    }
    
    \subfigure{
    \begin{minipage}[t]{1\linewidth}
    \centering
    \includegraphics[width=17cm]{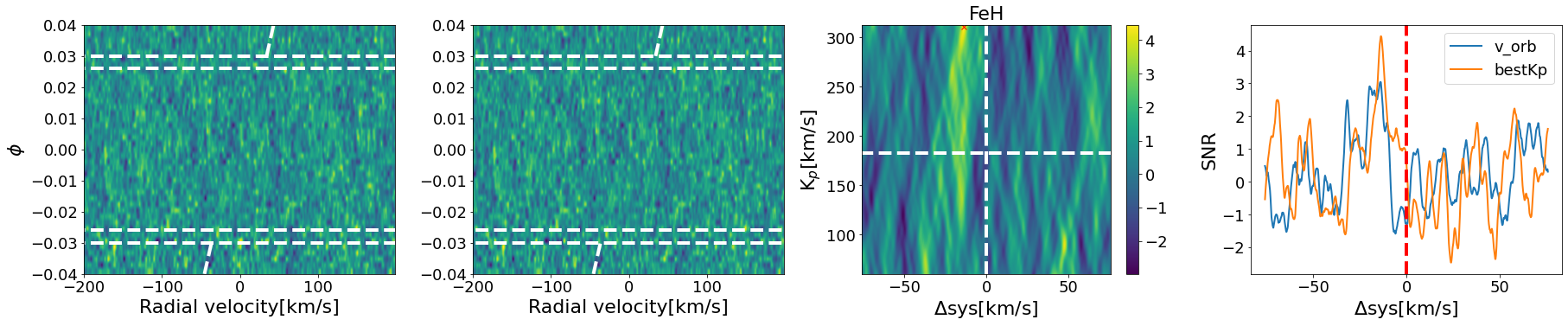}  
    \end{minipage}
    }
    
    \subfigure{
    \begin{minipage}[t]{1\linewidth}
    \centering
    \includegraphics[width=17cm]{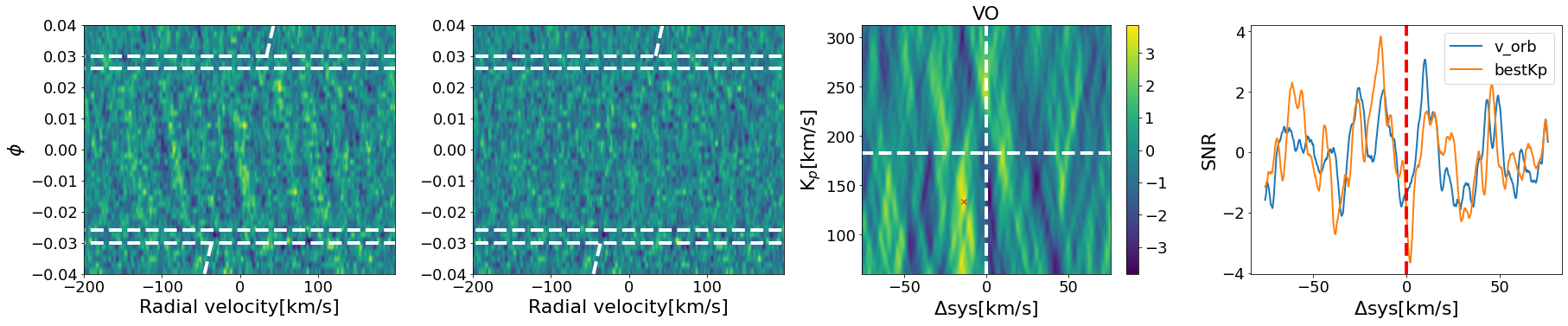}  
    \end{minipage}
    }

    \subfigure{
    \begin{minipage}[t]{1\linewidth}
    \centering
    \includegraphics[width=17cm]{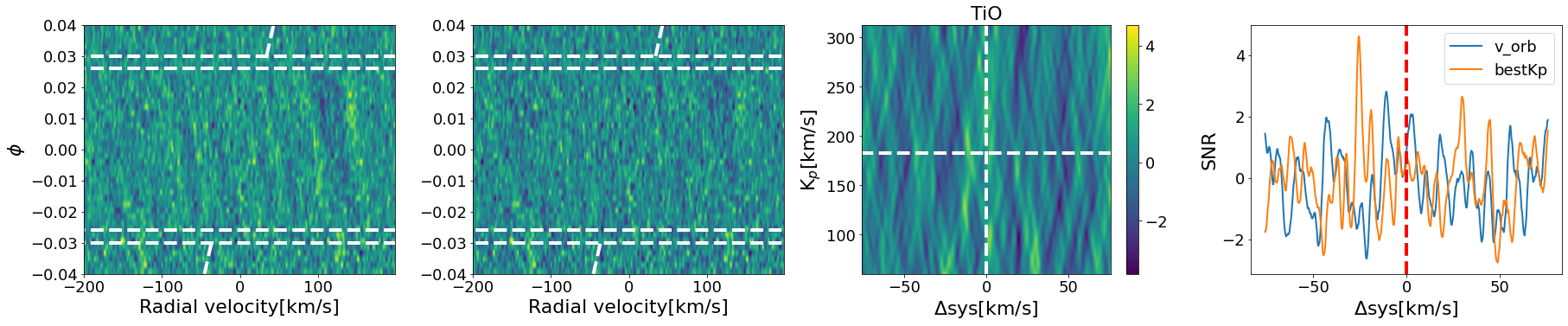}  
    \end{minipage}
    }

\caption{Same as Fig~\ref{fig:ccf_result_petit_1}: but for CaH, FeH, VO and TiO.}
    \label{fig:ccf_result_petit_2}
\end{figure*}

\begin{figure*}
    \centering
    
    \subfigure{
    \begin{minipage}[t]{1\linewidth}
    \centering
    \includegraphics[width=17cm]{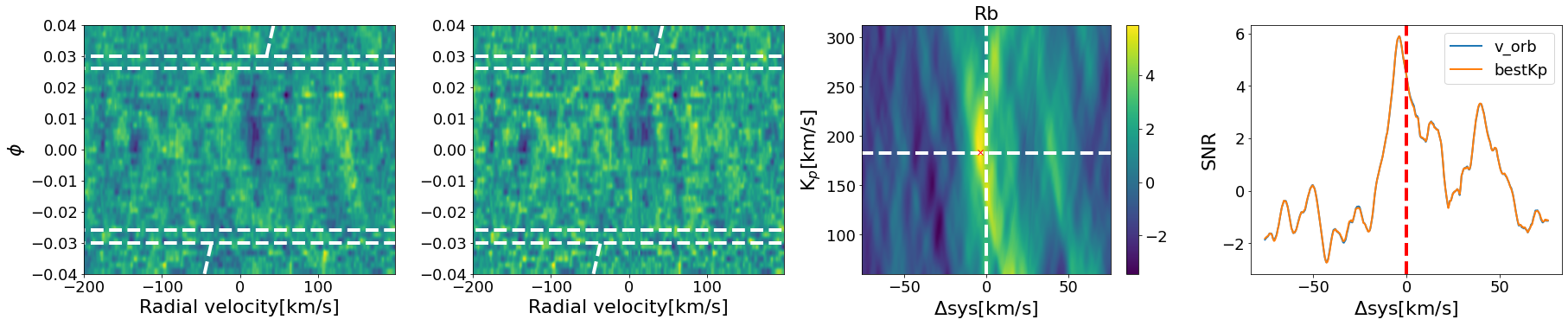} 
    \end{minipage}
    }
    
    \subfigure{
    \begin{minipage}[t]{1\linewidth}
    \centering
    \includegraphics[width=17cm]{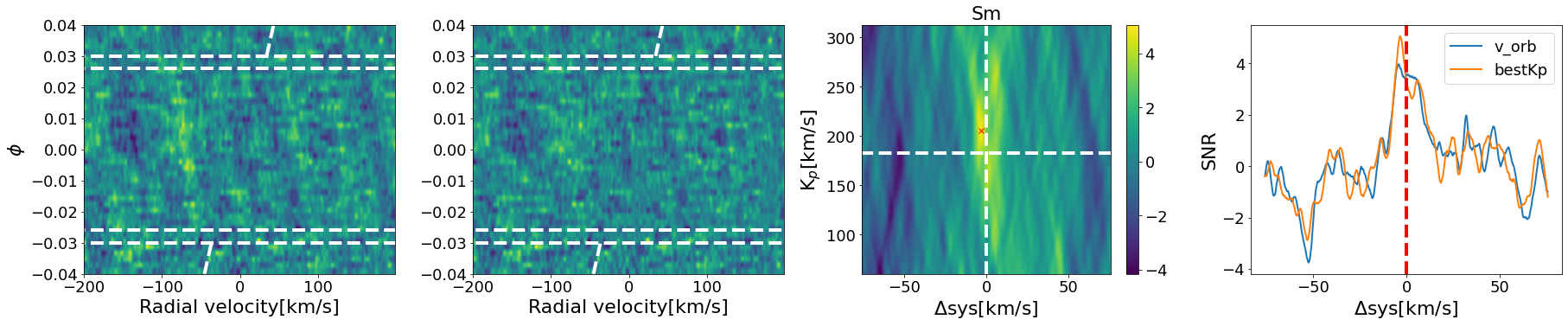}  
    \end{minipage}
    }
    
    \subfigure{
    \begin{minipage}[t]{1\linewidth}
    \centering
    \includegraphics[width=17cm]{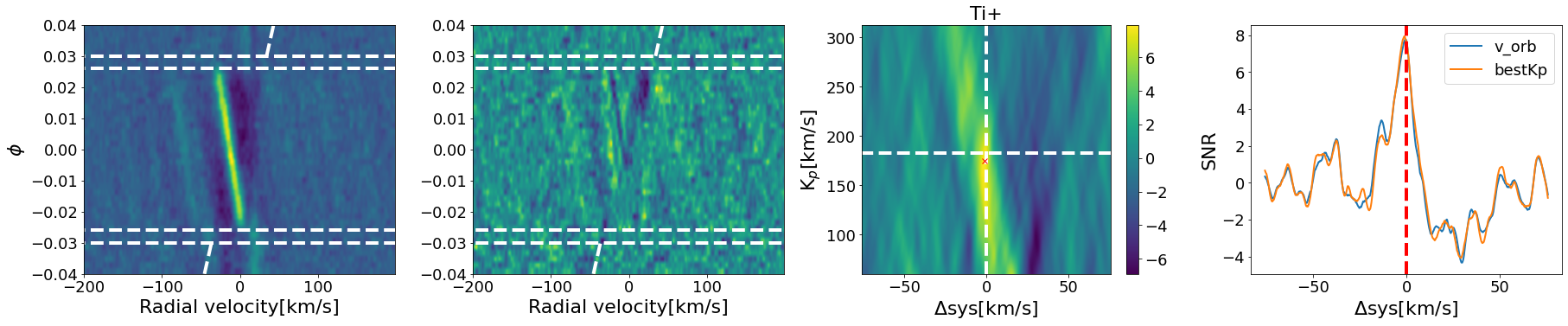}  
    \end{minipage}
    }
    
    \subfigure{
    \begin{minipage}[t]{1\linewidth}
    \centering
    \includegraphics[width=17cm]{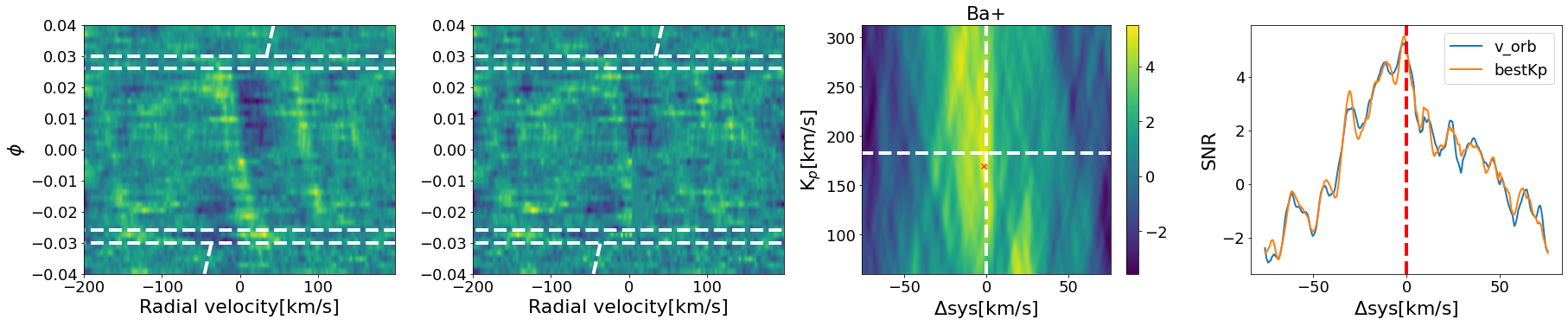}  
    \end{minipage}
    }

    \subfigure{
    \begin{minipage}[t]{1\linewidth}
    \centering
    \includegraphics[width=17cm]{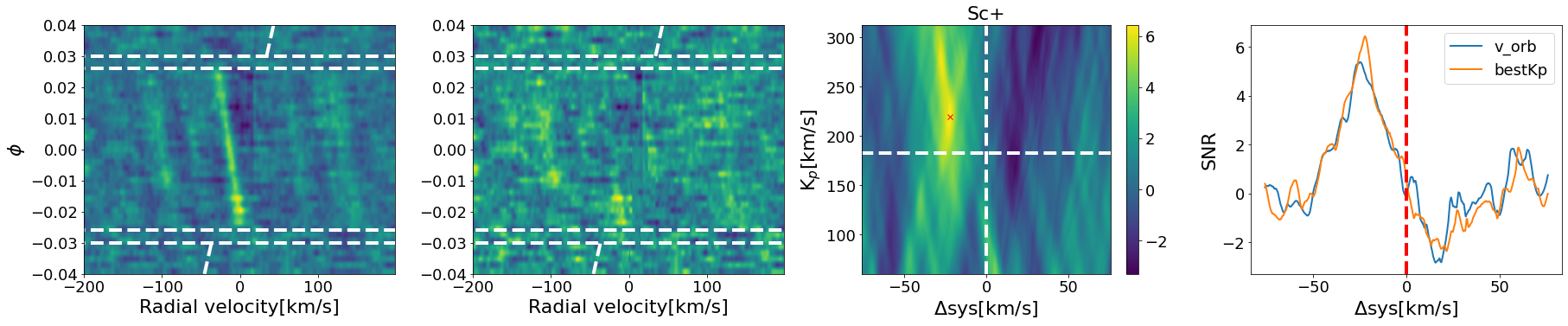}  
    \end{minipage}
    }
    
\caption{Same as Fig~\ref{fig:ccf_result_dace_1}: but for Rb, Sm, Ti+, Ba+ and Sc+.}
    \label{fig:ccf_result_dace_2}
\end{figure*}

As expected, we detect several heavier species in the atmosphere of MASCARA-4\,b, including Rb, \cc{Sm}, Ti+ and Ba+ with SNR$>5$ and reasonable offsets of $K_{\rm p}$ and $\Delta V_{\rm sys}$, as shown in Fig.~\ref{fig:ccf_result_dace_2} and Table~\ref{Gaussian fit}. This is the first time that the element Rb (Rubidium) is detected in the atmosphere of exopanet. Rb is an alkali metal similar to Na and K with atomic number 37, and is the most electropositive just after Caesium (Cs). Ba+ was firstly detected in two UHJs by \cite{Silva_2022}, and this work presents the third detection, giving solid support to the popular presence of Ba+ in UHJs. Ti+ was detected in KELT-9\,b~\citep{Hoeijmakers_2018} and WASP-189\,b~\citep{Prinoth_2022}, and was tentatively detected in WASP121\,b~\citep{Silva_2022} and WASP-33\,b~\citep{Cont_2022}, and is robustly detected in MASCARA-4\,b with high significance and small offsets. \cc{We also use the \texttt{pRT} to search for the molecules CaH, FeH, VO and TiO as shown in the Figure~\ref{fig:ccf_result_petit_2}. They all strongly absorb ultra-violet and visible light and would heat the upper atmosphere, leading to a rise in temperature with altitude, called as the thermal inversion ~\citep{Knutson_2008,Merritt_2020}. No signals from them are detected in this planet.}

In addition, we report tentative detection of \cc{Sc+}, with a peak SNR value about \cc{6.4 but deviating far from the estimated position}, with a relatively large $K_{\rm p}$ and $\Delta V_{\rm sys}$, as shown in Fig.~\ref{fig:ccf_result_dace_2} and Table~\ref{Gaussian fit}. The species \cc{Sc+} was detected previously in several UHJs, e.g., KELT-9\,b~\citep{Hoeijmakers_2018}. The common presence of those heavy ions (Ti+, Ba+, Sc+) at high altitudes in UHJ's implies that such kind of planets may experience atmospheric dynamics that we currently do not expect. From Fig.~\ref{fig:ccf_result_petit_1}, \ref{fig:ccf_result_dace_1}, \ref{fig:ccf_result_dace_2} and Table~\ref{Gaussian fit}, all the detected species excepting for Sc+ are blue-shifted by about \cc{0 to 10}\,km\,s$^{-1}$ as observed in the same target by \citet{Zhangyapeng_2022} and in WASP76\,b and WASP121\,b by \citet{Silva_2022}. Such phenomena is widely recognized as the result of the eastern wind across the terminator region of a tidally-locked planet.

Finally, it is quite interesting to note the \cc{new detection} of Sm (samarium), a lanthanide series element with atom number of 62. \cc{Sm is detected with peak SNR about 5.1 with velocity close to the planet $V_{\rm sys}$,} and is likely to a real detection. If confirmed, this will be the first lanthanide series element that has been detected in exoplanets, and is the heaviest element found so far. To summarize, all the Period 4 elements until Ni, plus several Period 5 and 6 metals have been detected in the atmospheres of UHJs. It is the right time to call for detailed atmospheric modeling to understand how these atoms and ions are populated at high altitudes. On the other hand, higher precision HRS studies should be further conducted on UHJs to reveal the presence of additional species and even some relatively abundant minor isotopes. Perhaps more importantly, retrieval analysis should be performed on the high-resolution transmission spectra to determine the abundance ratio of different elements, which when compared with those of their host stars, should shed light in the understanding of the formation and evolution of planets and their atmospheres. 

%---------------------------------------------------------------------------------------------------------------------------------%

%%%%%%%%%%%%%%%%%%%%%%%%%%%%%%%%%%%%%%%%%%%%%%%%
%%%%%%%%%%%%%%%%%%%%%%%%%%%%%%%%%%%%%%%%%%%%%%%

%%%%%%%%%%%%%%%%%%%%%%%%%%%%%%%%%%%%%%%%%%%%%%%%
%%%%%%%%%%%%%%%%%%%%%%%%%%%%%%%%%%%%%%%%%%%%%%%
%%%%%%%%%%%%%%%%%%%%%%%%%%%%%%%%%%%%%%%%%%%%%%%%
%%%%%%%%%%%%%%%%%%%%%%%%%%%%%%%%%%%%%%%%%%%%%%%

%% Also note that the akcnowlodgment environment does not support long amounts of text. If you have a lot of people and institutions to acknowledge, do not use this command. Instead, create a new \section{Acknowledgments}.

\section{Acknowledgments}
\begin{acknowledgments}
\cc{We thank the anonymous reviewer for their constructive comments.}
This research is supported by the National Natural Science Foundation of China grants No. 11988101, 42075123, 42005098, 62127901, 12273055, 11973048, 11927804, the National Key R\&D Program of China No.~2019YFA0405102, the Strategic Priority Research Program of Chinese Academy of Sciences, Grant No.~XDA15072113, the China Manned Space Project with NO. CMS-CSST-2021-B12. MZ, YQS are supported by the Chinese Academy of Sciences (CAS), through a grant to the CAS South America Center for Astronomy (CASSACA) in Santiago, Chile.
\end{acknowledgments}

\bibliography{sample631}{}
\bibliographystyle{aasjournal}



\end{document}